\documentclass[a4paper]{article}
\usepackage{graphicx}

\usepackage{amsmath}
\usepackage{amssymb}
\usepackage{latexsym}
\usepackage{epsfig}
\usepackage{amscd}
\usepackage{bbold,stmaryrd,bm}

\begin{document}
\title{Entanglement, decoherence and thermal relaxation in
exactly solvable models}

\author{Oleg Lychkovskiy\footnote{{\bf e-mail}: lychkovskiy@itep.ru}\\
\small{\em Institute for Theoretical and Experimental Physics} \\
\small{\em 117218, B.Cheremushkinskaya 25,
Moscow, Russia}
}

\maketitle

\newcommand{\be}{\begin{equation}}
\newcommand{\ee}{\end{equation}}

\newcommand{\la}{\langle}
\newcommand{\ra}{\rangle}

\newcommand{\ii}{\mathrm{i}}
\newcommand{\pp}{\mathbf{p}}
\newcommand{\ssigma}{\bm{\sigma}}

\newcommand{\cH}{{\cal H}}
\newcommand{\cS}{{\cal S}}
\newcommand{\cB}{{\cal B}}

\newcommand{\hPS}{ P^{\cal S}}
\newcommand{\hPE}{ P^{\cal E}}
\newcommand{\hH}{ H}
\newcommand{\hHB}{ H^{\cal B}}
\newcommand{\hHS}{ H^{\cal S}}
\newcommand{\hHSB}{ H^{\cal SB}}
\newcommand{\hV}{ V}
\newcommand{\hS}{ S}
\newcommand{\psiS}{\psi^{\cal S}}
\newcommand{\psiB}{\psi^{\cal B}}

\newcommand{\dens}{ \rho}
\newcommand{\densS}{ \rho^{\cal S}}
\newcommand{\densB}{ \rho^{\cal B}}
\newcommand{\meandens}{{ \bar \rho}}
\newcommand{\meandensS}{\overline{{ \rho}^{\cal S}}}
\newcommand{\meanavdensS}{\langle \overline{{ \rho}^{\cal S}}\rangle_{\cal B}}
\newcommand{\tr}{\rm tr}

\begin{abstract}
Exactly solvable models provide an opportunity to study different aspects of reduced quantum dynamics in detail. We consider the reduced dynamics of a single spin in finite $XX$ and $XY$ spin $1/2$ chains. First we introduce a general expression describing the evolution of the reduced density matrix. This expression proves to be tractable when the combined closed system (i.e. open system plus environment) is integrable. Then we focus on comparing decoherence and thermalization timescales in the $XX$ chain. We find that for a single spin  these timescales are comparable, in contrast to what should be expected for a macroscopic body. This indicates that the process of quantum relaxation of a system with few accessible states can not be separated in two distinct stages -- decoherence and thermalization. Finally, we turn to finite-size effects in the time evolution of a single spin in the $XY$ chain. We observe  three consecutive stages of the evolution: regular evolution, partial revivals, irregular (apparently chaotic) evolution. The duration of the regular stage is proportional to the number of spins in the chain. We observe a "quiet and cold period" in the end of the regular stage, which breaks up abruptly at some threshold time.
\end{abstract}


\section{Introduction}

In the present contribution several results and observations are presented which concern the dynamics of exactly solvable models and especially one-dimensional spin chains. The integrability allows to obtain rigorous results, however we believe that the applicability of general conclusions based on these results is not restricted by integrable models only. The paper is organized as follows. In the next section the quantum Liouville equation is rewritten in the form which may be tractable when the model is exactly solvable. In section 3 a sketch of the $XY$ model is given. In section 4 a comparison of decoherence and thermalization timescales in the isotropic $XY$ (i.e. $XX$) model is carried out. Section 5 is devoted to the dynamical regular-to-chaotic transition in the $XY$ chain. In section 6 we focus on the quiet and cold period in evolution of a spin in the $XY$ chain which takes place in the end of the regular evolution. The details of calculations are presented in two appendices.

\section{Reduced dynamics of a spin: general expression}

Consider a quantum system (described by a Hilbert space $\cS$), which interacts with a quantum environment or bath (described by a Hilbert space $\cB$). The composite system with Hilbert space $\cH=\cS \otimes \cB$ is considered to be closed. Its density matrix evolves according to
\be
\dens(t)=\exp(-i \hH t)\dens(0) \exp(i \hH t).
\ee
The Hamiltonian may be decomposed as
\be
\hH=\hHS+\hHB+\hHSB,
\ee
where $\hHS$ and $\hHB$ are self-Hamiltonians of the system and the bath correspondingly, and $\hHSB$ is an interaction Hamiltonian.
Here and in what follows the usage of superscripts and subscripts $\cS,~\cB$ is believed to be self-explanatory.


The spectral decomposition of the total Hamiltonian reads
\be
\hH=\sum_{n=1}^d E_n |\Psi_n\ra\la\Psi_n|,
\ee
where $\Psi_n$ are the eigenvectors and $d\equiv\dim \cH =d_{\cS}d_{\cB}$.

The time evolution of $\densS(t)$ reads
\be
\densS(t)=\sum_{n=1}^d \sum_{m=1}^d  \la \Psi_n |\dens(0)| \Psi_m \ra e^{-i (E_n-E_m)t} \densS_{nm},
\ee
where the $d_{\cS}\times d_{\cS}$ matrices
\be
\densS_{nm} \equiv \tr_{\cB}|\Psi_n\ra\la \Psi_m|
\ee
are introduced (not to be confused with matrix elements!). Evidently, these matrices encode the dynamics of the open system $\cS,$ while the matrix elements $\la \Psi_n |\dens(0)| \Psi_m \ra$ describe the initial conditions.

In the case when the system $\cS$ is represented by a single spin, its reduced density matrix $\densS(t)$ and matrices $\densS_{nm}$ may be parameterized by polarization vectors:
\be \densS(t)=(1+\pp(t)\ssigma)/2, ~~~ \densS_{nm}=(\delta_{nm}+\pp_{nm}\ssigma)/2.
\ee
The polarization vector of a spin $\pp(t)$ belongs to a unit sphere which is known as the Bloch sphere. The length of the polarization vector equals 1 for a pure state and is less than 1 for a mixed state. Its time evolution is given by
\be\label{p(t)}
\pp(t)=\sum\limits_{n=1}^d \sum\limits_{m=1}^d  \la \Psi_n |\dens(0)| \Psi_m \ra e^{-i (E_n-E_m)t} \pp_{nm}.
\ee
As a rule this expression is not very useful in practical calculations (however, it is used to prove a necessary condition for thermalization of a spin coupled to a bath, see  ref. \cite{Lychkovskiy}). However, when the model is integrable, one has a chance to calculate $\pp_{nm}$ explicitly. If this is done and if the initial condition is of some simple form (e.g., the bath is in a thermal state), then eq.(\ref{p(t)}) may appear to be tractable and allow to study the reduced evolution of a spin in detail. We will demonstrate this is the case of integrable one-dimensional spin chain -- the $XY$ chain. Before we turn to this specific model let us give an example of initial conditions that simplify the problem greatly. These conditions correspond to the uncorrelated system and environment, when the system is in arbitrary pure state, and the environment is in thermal state with infinite temperature:
\be\label{infinite T initial conditions}
\dens(0)= 2^{-N} (1+\pp_0\ssigma)\otimes  1^{\cB}.
\ee
In this case eq.(\ref{p(t)}) reads
\be\label{p(t) at beta=0}
\pp(t)=2^{-N} \sum\limits_{n=1}^d \sum\limits_{m=1}^d  (\pp_0,\pp_{mn}) e^{-i (E_n-E_m)t} \pp_{nm}.
\ee

%
%

\section{XY model on a circle}

Consider a chain of $N$ coupled spins $1/2$ with the following Hamiltonian:
\be \hH =\frac{\kappa}4  \sum_{j=1}^N ((1+\gamma)\sigma_j^x \sigma_{j+1}^x+ (1-\gamma) \sigma_j^y \sigma_{j+1}^y)  + \frac{h}2  \sum_{j=1}^N \sigma_j^z.
\ee
Here the index $N+1$ is identified with $1,$ and $N$ is supposed to be even. The following parameters enter the Hamiltonian:
\begin{itemize}
\item $\kappa$ -- coupling constant,
\item $\gamma$ -- anisotropy parameter,
\item $h$ -- magnetic field.
\end{itemize}
We consider the case of weak and moderate coupling: $h>\kappa.$

For the purposes of the present study we regard the first spin as the system $\cS,$ and other $(N-1)$ spins -- as the environment $\cB.$

Finite $XY$ model is "almost diagonalizable" through the sequential Jordan-Wigner, Fourier and Bogolyubov transformations (see e.g. \cite{Izergin}\cite{Pasquale}). An important property of the $XY$ Hamiltonian is that it commutes with the parity operator $\Pi\equiv \prod_{j=1}^N \sigma^z:$
\be
[\Pi,H]=0.
\ee
The diagonalization leads to the ``almost free fermion form'' of the Hamiltonian:

\be
H=P^{\rm odd} \sum_{q\in X_{\rm odd}} E_q (c_{q}^+ c_{q}-\frac12) + P^{\rm ev} \sum_{q\in X_{\rm ev}} E_q (c_{q}^+ c_{q}-\frac12),
\ee
where
\be
X_{\rm odd}=\{-\frac{N}2+1,-\frac{N}2+2,...,\frac{N}2\},
~~~X_{\rm ev}=\{-\frac{N}2+\frac12,-\frac{N}2+\frac32,...,\frac{N}2-\frac12\},
\ee
 $\{ c_{q},~ q\in X_{\rm odd} \}$ and $\{ c_{q},~q\in X_{\rm ev} \}$ are two sets of fermion operators (note, however, that the operators from different sets do not satisfy anticommutation relations, see eq.(\ref{c anticommutation})), $P^{\rm odd}$ and $P^{\rm ev}$ are parity projectors,
\be
P^{\rm ev}\equiv (1+\Pi)/2,~~~~ P^{\rm odd}\equiv (1-\Pi)/2
\ee
and fermion energy is defined as
\be
 E_q=\sqrt{(h-\kappa\cos(2\pi q/N))^2+(\gamma\kappa \sin(2\pi q/N))^2}.
\ee

The details of the $XY$ spin chain diagonalization are presented in the appendix A.

\section{Decoherence versus thermalization timescales}

\begin{figure}[t]
\begin{center}
$
\begin{array}{cc}
 \includegraphics[width=0.5\textwidth]{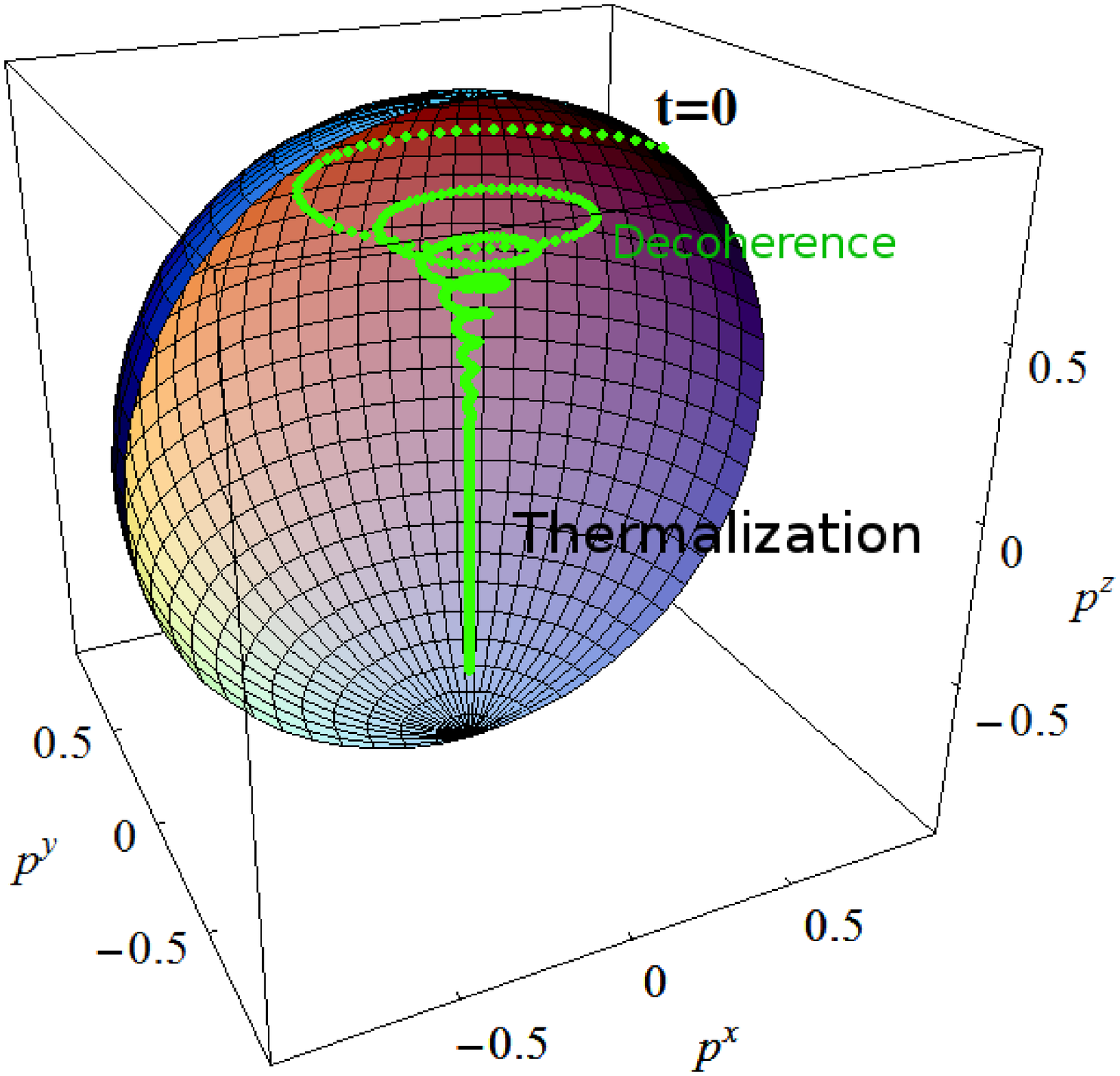}  & \includegraphics[width=0.5\textwidth]{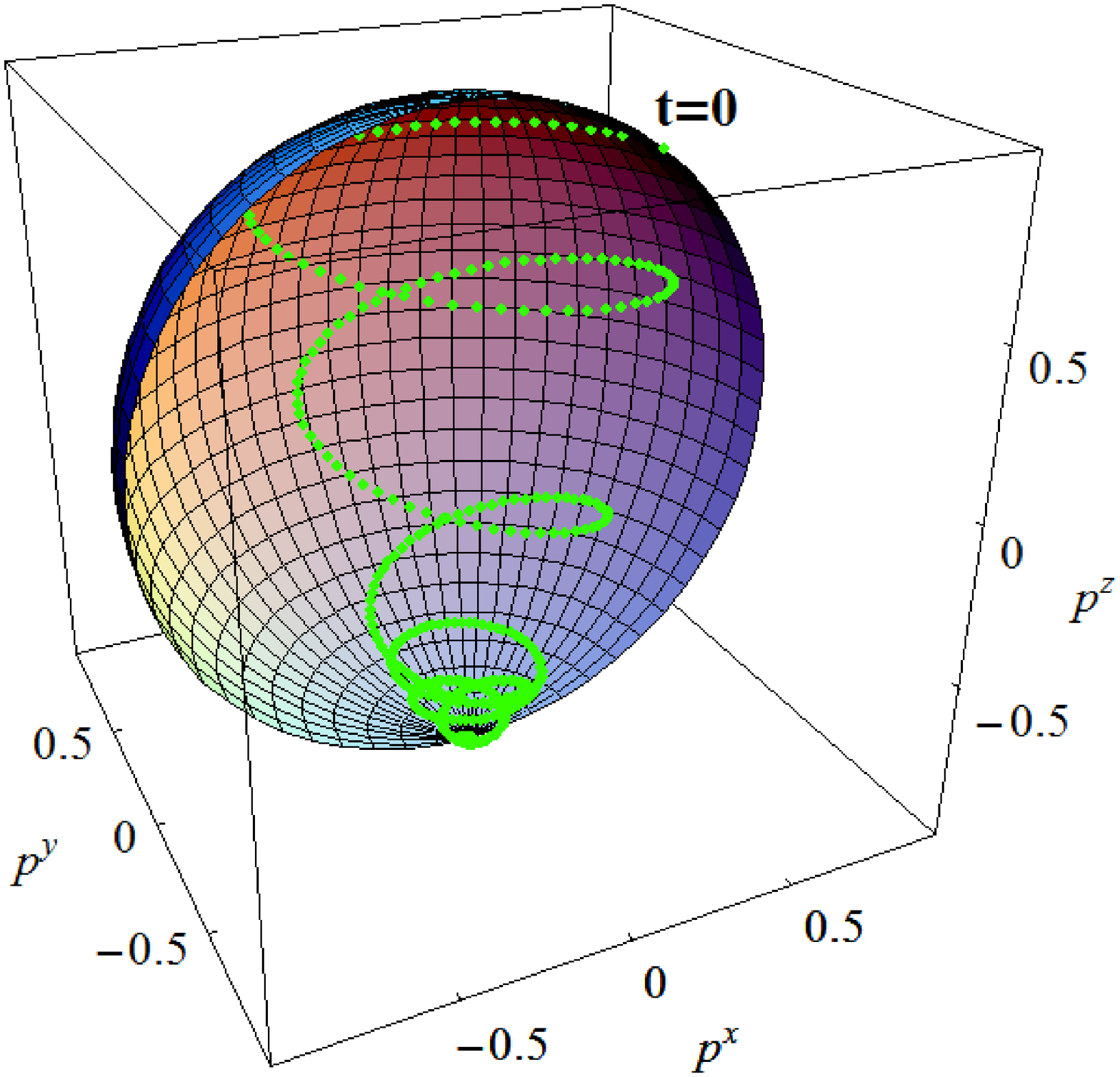}\\
\end{array}
$
\end{center}
\caption{\label{decoherence vs thermalization} Patterns of the spin polarization evolution $\pp(t)$. Left: what may be expected for the relaxation of a spin in the analogy to the relaxation of macroscopic bodies (``artist's view''). Right: real relaxation of a spin in the $XX$ chain for $h=10\kappa$ (plot). The points correspond to the values of $\pp(t)$ from $t=0$ to $t=4\kappa^{-1}$ with the step $0.01\kappa^{-1}.$ The initial value reads $\pp(0)=(0.6,0,0.8).$ See the text for other details.
}
\end{figure}


Decoherence is a process of the dynamical vanishing of non-diagonal entries of the reduced density matrix  $\densS(t)$ written in the pointer basis. Thermalization is a process of the energy exchange between the system and the environment leading to thermal equilibrium. What is the relation between the timescales of these processes? Evidently, for macroscopic bodies the decoherence timescale is much less than the thermalization timescale. Namely, decoherence timescale should be {\it microscopic}, in order to ensure the fast quantum-to-classical transition as it is required by our everyday experience. In contrast, the thermalization timescale is usually  {\it macroscopic}. In fact, this hierarchy of the timescales is the basis for the very distinction between the decoherence and the thermalization: the former emerges as a first short stage of quantum relaxation, and the latter -- as a succeeding long stage.

Is the above reasoning applicable to small quantum systems, in particular, to a single spin~$1/2$? Very often one implicitly assumes that it is. However, below we present a counterexample: the decoherence and thermalization timescales of a spin in the $XY$ spin chain appear to be of the same order.

The easiest way to show this is to consider a special case with zero anisotropy ($\gamma=0,$ $XX$ chain) and small coupling ($\kappa \ll h$). The latter condition ensures that the pointer basis is the eigen basis of the self-Hamiltonian $\frac{h}2 \sigma_1^z$ \cite{Paz and Zurek} in the sense that $\densS(t)$ is approximately diagonal in this basis for almost all times \cite{Gogolin}. Let us take the initial state to be
\be \dens(0)= (1+\pp_0\ssigma_1)/2\otimes |\downarrow_2\downarrow_3...\downarrow_N\ra\la\downarrow_2\downarrow_3...\downarrow_N|,
\ee
which corresponds to the environment at zero temperature and arbitrary polarization $\pp_0$ of the first spin. We are interested in the evolution of the central spin polarization $\pp(t)$. In the analogy to the relaxation in macroworld one could expect the following relaxation pattern (see the left picture on Fig. 1). First $p_x$ and $p_y$ would vanish abruptly (or, equivalently, $\densS(t)$ would become approximately diagonal in the pointer basis $\{ |\downarrow_1\ra, ~|\uparrow_1\ra \}$), which would constitute decoherence. In contrast, $p_z$ would decrease  gradually revealing the energy flow from the first spin (i.e. system) to cold environment. This pattern would allow to unambiguously separate decoherence from thermalization, as is evident from Fig. 1, left.

However the described above pattern is not realized in reality. Instead decoherence and thermalization proceed simultaneously, see Fig. 1, right.
To be more exact, the process of relaxation can not be separated in time in two different stages -- decoherence and thermalization. This can be seen from the approximate analytical solution which is valid for $ 0\leq t< N/\kappa$  \cite{Niemeijer}:
\be \begin{array}{cl}
p^x= & J_0(\kappa t) (p_0^x \cos ht - p_0^y \sin ht),\\
p^y= & J_0(\kappa t) (p_0^x \sin ht + p_0^y \cos ht),\\
p^z= & -1+(1+p_0^z)J_0^2(\kappa t),
\end{array}
\ee
where $J_0$ is the Bessel function.

One can see that the relaxation of all components of $\pp(t)$ occurs at a common timescale $\sim \kappa^{-1}.$ In a system which consists of a single spin there is no large dimensionless parameter which could be responsible for the hierarchy of timescales. This reasoning suggests that the absence of such hierarchy is a general property of quantum systems with small number of accessible states rather than a peculiar feature attributed to integrability of the $XY$ chain. However further work is necessary to show this with more rigor.

\section{Finite-size effects in $XY$ chain}

Integrable spin chains are often studied in the $N\rightarrow\infty$ approximation. Although this allows to use the powerful machinery of Quantum Field Theory, some interesting features of the evolution may disappear in this approximation. The present section is devoted to one of such features -- the regular-to-chaotic transition during the evolution of $\pp(t).$ A similar effect in a somewhat different setting is observed (but not discussed in detail) in \cite{Mossel}.

In what follows we take the initial state (\ref{infinite T initial conditions}) which corresponds to the environment at $T=\infty$. As was already discussed such choice of the initial conditions simplifies the problem. However one should be cautious when interpreting the infinite temperature state $2^{-N+1}1^{\cB}$. Indeed, consider a thermal state with some high temperature $T.$ It is well approximated by $2^{-N+1}1^{\cB}$ whenever $T \gg N h $ but not just  $T \gg h.$ Evidently the former condition (in contrast to the latter one) becomes more stringent with the growth of $N$ and appears to be impracticable in the limit  $N\rightarrow \infty.$ However this subtlety is not important in our discussion: the qualitative picture of the regular-to-chaotic transition does not depend on the temperature of the environment.

\begin{figure}[t]
\begin{center}
\includegraphics[width=1\textwidth]{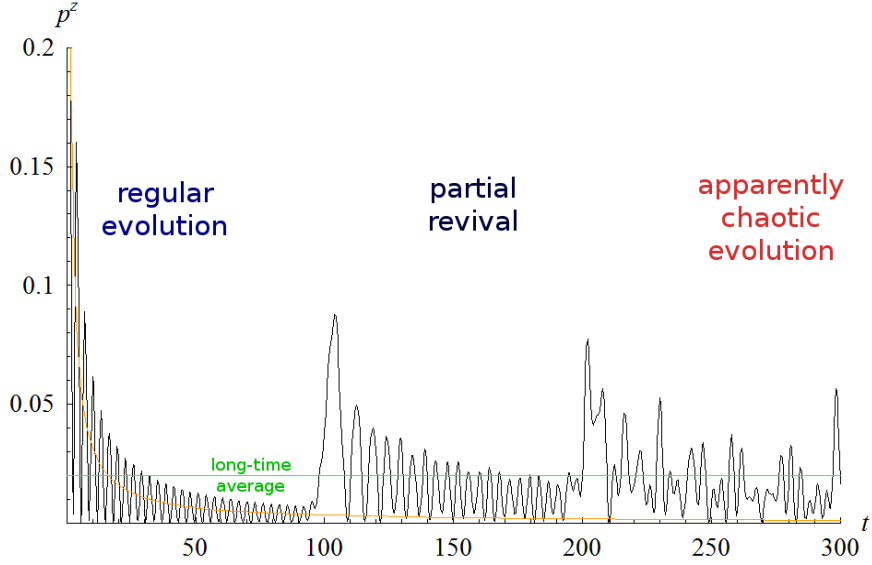}
\end{center}
\caption{\label{regular2chaotic} Evolution of $p^z(t)$ for $\gamma=0,~ h=5, ~\kappa=1, ~N=100.$  }
\end{figure}

We focus our attention on the evolution of the $z$-component of polarization. Using the explicit expressions for the eigenstates of the $XY$ model one gets from eq.(\ref{p(t) at beta=0}) the following {\it exact} formula (the details of calculations see in appendix~B):
\be\label{pz(t) at beta=0}
p^z(t) = \frac12 p_0^z (A_{\rm odd}^2+A_{\rm ev}^2+B_{\rm odd}^2+B_{\rm ev}^2),
\ee
where
\be\label{pz(t) at beta=0, definitions}
\begin{array}{rclcrcl}
A_{\rm ev}(t) & = &N^{-1}\sum\limits_{q\in X_{\rm ev}}\cos E_q  t,  & & A_{\rm odd}(t) & = & N^{-1}\sum\limits_{q\in X_{\rm odd}}\cos E_q  t, \\
B_{\rm ev}(t) & = & N^{-1}\sum\limits_{q\in X_{\rm ev}}\cos\theta_q \sin E_q  t, & & B_{\rm odd}(t) & = & N^{-1}\sum\limits_{q\in X_{\rm odd}}\cos\theta_q \sin E_q  t,
\end{array}
\ee
\be
\cos \theta_q= \frac{h-\kappa\cos(2\pi q/N)}{E_q}.
\ee

The plot of $p^z(t)$ for $\gamma=0$ ($XX$ chain), $h=5, ~\kappa=1,~ N=100$ is presented on Fig. 2. One can distinguish three successive stages of evolution:
\begin{enumerate}
\item $0\leq \kappa t \lesssim N:$ regular evolution. At this stage $p^z(t)$ is well described by the approximate expression \cite{Niemeijer}
\be\label{pz(t) regular}
p^z(t) \simeq  p_0^z J_0^2(\kappa t).
\ee
\item $ N \lesssim \kappa t \lesssim {\rm few}~ N:$ partial revivals.
\item $ \kappa t \gtrsim {\rm few}~ N:$ apparently chaotic evolution.
\end{enumerate}

As the duration of the regular stage is proportional to $N,$ it is the only stage which may be catched by the $N\rightarrow\infty$ approximation. Evidently this stage ends up when the perturbation initially localized in the first site of the chain propagates through the circle and returns back to the first site. However the abruptness of the deviation of $p^z(t)$ from the regular expression (\ref{pz(t) regular}) at some threshold time calls for additional explanation, which will be the purpose of the future work.

The second stage has a well-defined starting point, but its termination point is not well-defined. One can clearly see one partial revival at $N \lesssim \kappa t \lesssim 2N$ in Fig. 2, but the succeeding evolution can not be characterized that unambiguously. In fact the second stage gradually changes to the third one -- apparently chaotic. Note that due to the Poincare recurrence theorem the revivals should interrupt the chaotic evolution at long times.

On Fig. 3 one can see the plots for $p^z(t)$ for various anisotropies and magnetic fields. Evidently large anisotropy suppresses the oscillations, while  large magnetic field  amplifies them.

\begin{figure}[t]
\begin{center}
$
\begin{array}{cc}
  \includegraphics[width=0.5\textwidth]{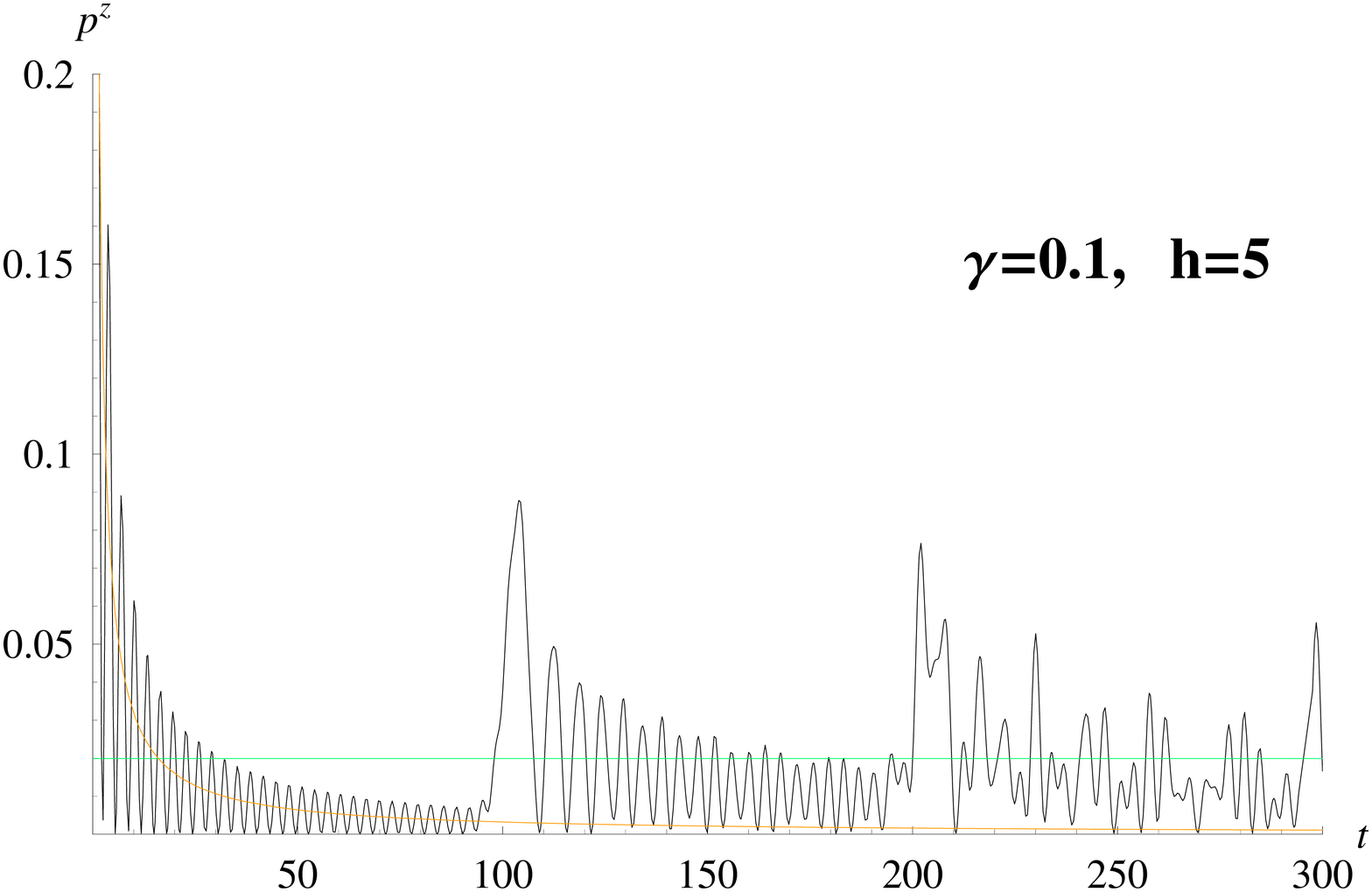}  & \includegraphics[width=0.5\textwidth]{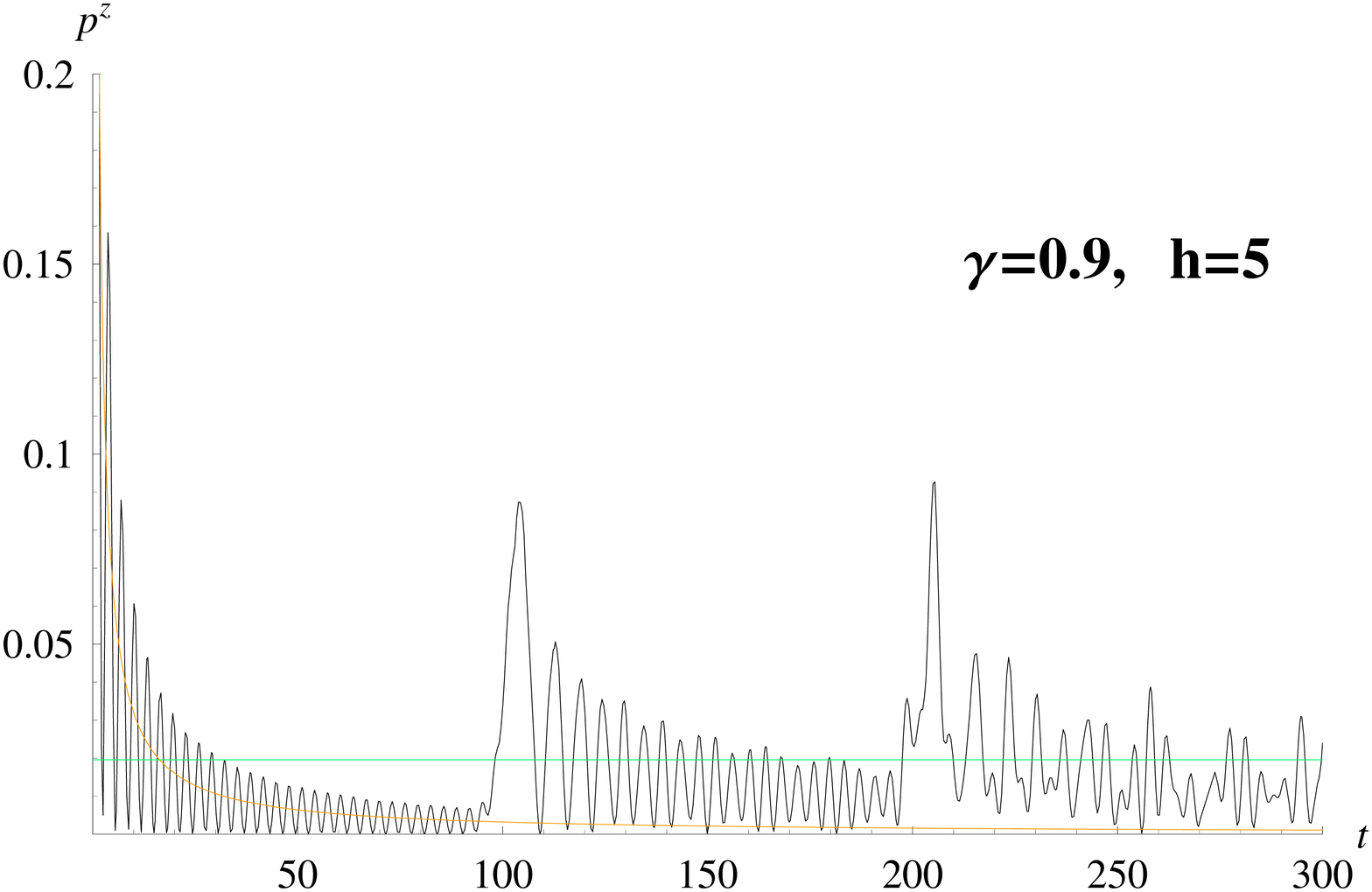}\\
  \includegraphics[width=0.5\textwidth]{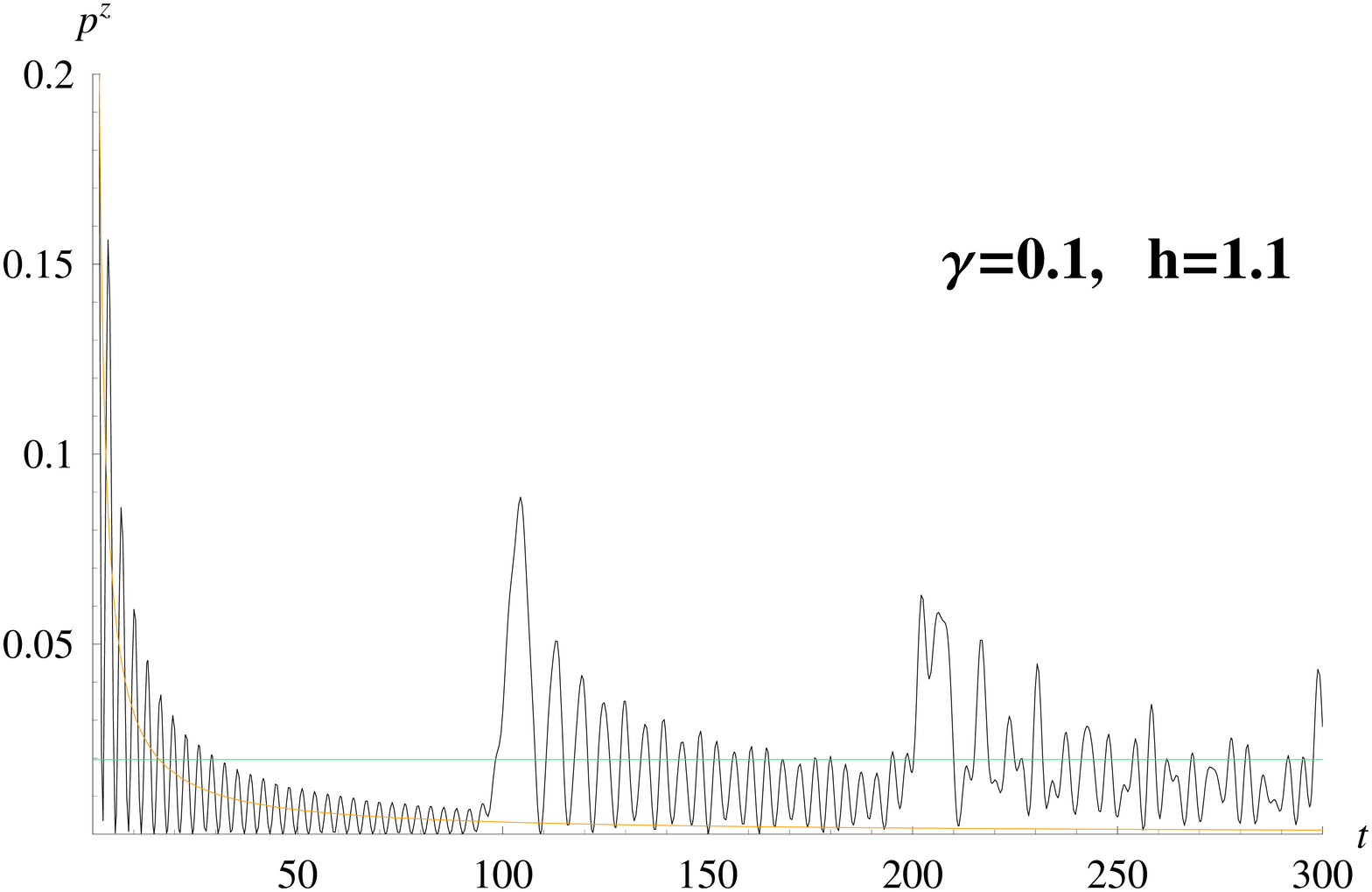}  & \includegraphics[width=0.5\textwidth]{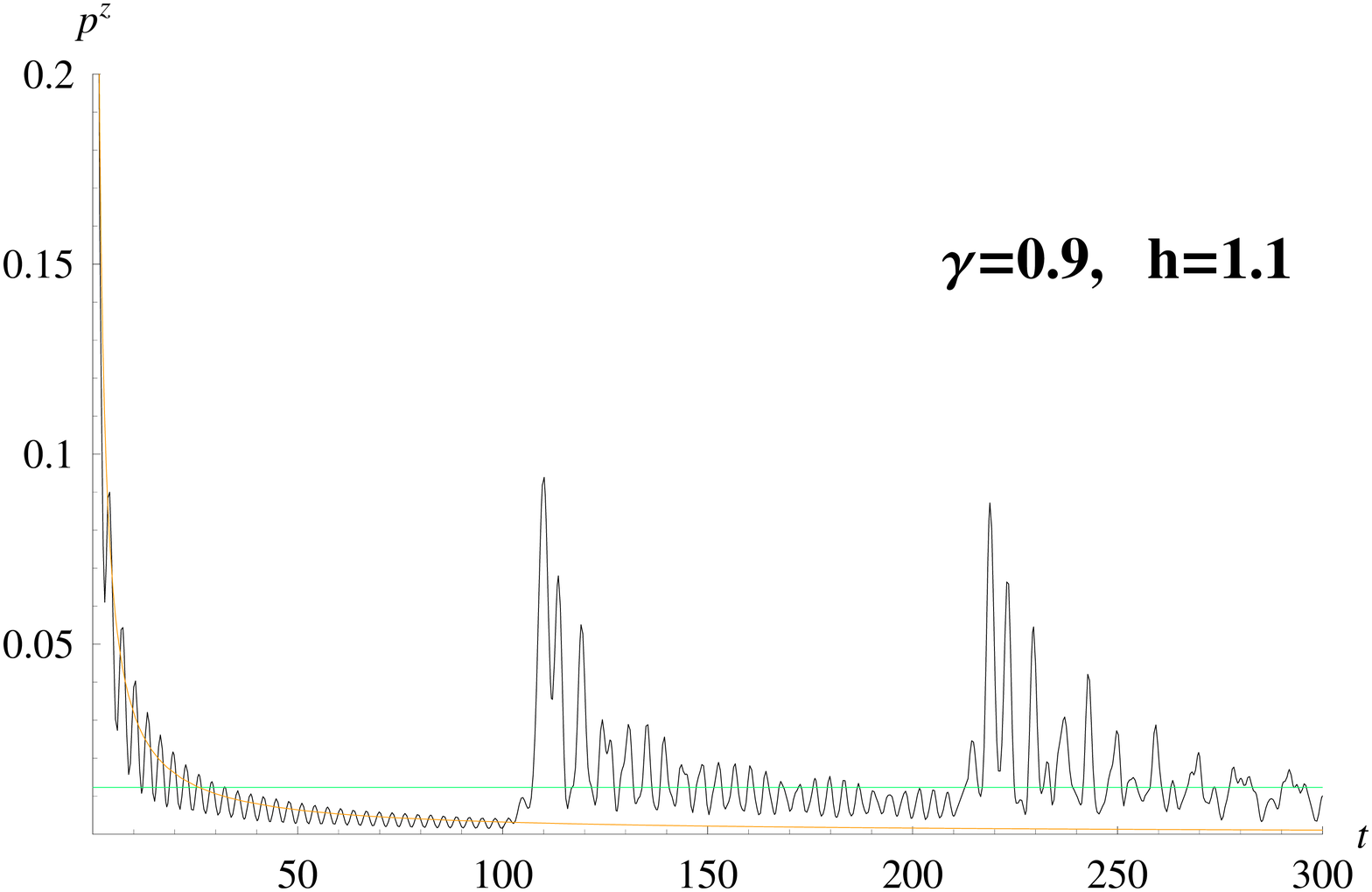}
\end{array}
$
\end{center}
\caption{\label{various parameters} Evolution of $p^z(t)$ for various anisotropies and magnetic fields. $\kappa=1,~ N=100.$}
\end{figure}

\section{Quiet and cold period}
The asymptotic behavior of the Bessel function is given by
\be
J_0(x) \simeq \sqrt{2/\pi x} \cos(x-\pi/4).
\ee
Therefore, according to eq.(\ref{pz(t) regular}), during the regular stage the smoothed behavior of $p^z(t)$ in the $XX$ chain is described by
\be
\tilde p^z(t) = \frac{p_0^z}{\pi \kappa t}.
\ee
At the same time the long-time average of $p^z(t)$ is given by
\be
\overline p^z \simeq 2p_0^z/N.
\ee
Note that at times $N (2\pi\kappa)^{-1}\lesssim  t \lesssim N\kappa^{-1}$ the smoothed $p^z(t)$ is {\it less} than the average value  $\overline p^z.$
This means that during this time interval the energy of the central spin is less than its long-time average energy. Also the oscillations of $p^z(t)$ are suppressed at this time. Thus we observe the quiet and cold period which precedes the onset of chaotic evolution. Such "calm before the storm" may be of practical importance in the context of quantum information processing.

\section{Acknowledgements}
I am grateful to E. Safonov for useful discussions.
I acknowledge the partial support from grants NSh-4172.2010.2, RFBR-11-02-00778, RFBR-10-02-01398 and from the Ministry of Education and Science of the Russian Federation under contracts N$^{\underline{\rm o}}$N$^{\underline{\rm o}}$ 02.740.11.5158, 02.740.11.0239.

\section*{Appendix A: diagonalization of finite $XY$ spin chain}
The diagonalization of the $XY$ spin chain was studied in many works starting from the pioneering paper \cite{Lieb}. The early papers focused on the thermodynamic $N\rightarrow \infty$ limit. The finite size $XY$ chain is discussed in later works, see e.g. \cite{Izergin}\cite{Pasquale}. Here we present the details of diagonalization for the sake of completeness.

\subsection*{A.1~~~ $H$ in terms of $\sigma_j^\pm$}
We define the operators $\sigma_j^\pm$ in a usual way,
\be
\sigma_j^+=\frac12 (\sigma_j^x+i \sigma_j^y),~~~\sigma_j^+=\frac12 (\sigma_j^x-i \sigma_j^y).
\ee
These operators are neither Bose nor Fermi operators:
\be
\sigma_j^+ \sigma_j^- + \sigma_j^- \sigma_j^+=1,
\ee
\be
\sigma_j^+ \sigma_n^- = \sigma_n^- \sigma_j^+ {\rm~~~for~~~}j\neq n.
\ee
The following simple equalities prove to be useful:
\be
\sigma^z=2\sigma^+ \sigma^--1=-2\sigma^- \sigma^+ + 1
\ee
\be
\sigma^z\sigma^+=-\sigma^+\sigma^z=\sigma^+,~~~\sigma^z\sigma^-=-\sigma^-\sigma^z=-\sigma^-
\ee
The Hamiltonian may be rewritten in terms of $\sigma_j^\pm$ as follows:
\be
H=H_0+H_\gamma+H_h
\ee
with
\be
H_0 =\frac\kappa2 \sum_{j=1}^N (\sigma_j^+ \sigma_{j+1}^-+ \sigma_j^- \sigma_{j+1}^+),
\ee
\be
\hH_\gamma =\frac{\kappa\gamma}2 \sum_{j=1}^N (\sigma_j^+ \sigma_{j+1}^+ + \sigma_j^- \sigma_{j+1}^-),
\ee
\be
\hH_h =h \sum_{j=1}^N \sigma_j^+ \sigma_j^- -Nh/2.
\ee

\subsection*{A.2~~~ Jordan-Wigner transformation}
Define the operators
\be
\Pi_n \equiv \prod_{j=1}^n \sigma^z_j.
\ee
Evidently, $\Pi_N$ coincides with the parity operator $\Pi$ defined in section 3.

Define Fermi operators $a^-_j,~a^+_j$ as follows
\be
a^-_j \equiv \sigma_j^- \Pi_{j-1} =\Pi_{j-1} \sigma_j^-,~~~ a^+_j \equiv \sigma_j^+ \Pi_{j-1} =\Pi_{j-1} \sigma_j^+
\ee
This implies
\be
\sigma_j^-= a^-_j  \Pi_{j-1},~~~\sigma_j^+= a^+_j  \Pi_{j-1},
\ee

\be
\{a_m^+,a^-_n\}=\delta_{mn},~~~\{a_m^+,a^+_n\}=\{a_m^-,a^-_n\}=0,
\ee

\be
\sigma^z_j=2a^+_j a^-_j -1=-2a_j^- a_j^+ + 1.
\ee
The Hamiltonian takes the form (note that now the ordering of $a_j^\pm,~a_{j+1}^\pm$ is important; also note the change of the total sign):

\be
H_0 =- \frac\kappa2[ \sum_{j=1}^N (a_j^+ a_{j+1}^- + a_{j+1}^+a_j^-) - (1+\Pi)(a_N^+ a_1^- + a_1^+a_N^-)].
\ee

\be
\hH_\gamma =- \frac{\kappa\gamma}2[ \sum_{j=1}^N (a_j^+ a_{j+1}^+ + a_{j+1}^-a_j^-) - (1+\Pi)(a_N^+ a_1^+ + a_1^-a_N^-)].
\ee

\be
\hH_h =h \sum_{j=1}^N a_j^+ a_j^- -Nh/2.
\ee

\subsection*{A.3~~~ Fourier transformation}

Define for arbitrary real $q$
\be
b_q^-\equiv \frac{e^{i \pi/4}}{\sqrt N}\sum_{n=1}^N e^{-2 \pi i q (n-1)/N}a_n^-,
\ee
\be
b_q^+\equiv \frac{e^{-i \pi/4}}{\sqrt N}\sum_{n=1}^N e^{2 \pi i q (n-1)/N}a_n^+,
\ee
Then
\be
\{b_k^+,b^+_q\}=\{b_k^-,b^-_q\}=0,
\ee
\be\label{c anticommutation}
\{b_k^+,b^-_q\}=\frac1N\frac{1-e^{2\pi i (k-q)}}{1-e^{2\pi i (k-q)/N}}.
\ee
In particular, if one takes
\be
q=-\frac{N}2+1,~-\frac{N}2+2,...,\frac{N}2 ~~~~~~ (X_{\rm odd})
\ee
or
\be
q=\frac{N}2+\frac12,~-\frac{N}2+\frac32,...,\frac{N}2-\frac12 ~~~~~~ (X_{\rm ev})
\ee
then the set of $b_q^-$ is the set of Fermi annihilation operators.

The Hamiltonian may be written in terms of $b_q^\pm$ as follows:
\be
H=H^{\rm odd}P^{\rm odd}+H^{\rm ev}P^{\rm ev}
\ee
with
\be
P^{\rm odd}\equiv(1-\Pi)/2,~~~P^{\rm ev}\equiv(1+\Pi)/2,
\ee
and

\be
H_0^{\rm odd ~(ev)}=-\kappa \sum_{q\in X_{\rm odd ~(ev)}}~ \cos (2\pi q/N) b^+_q b^-_q,
\ee

\be
H_\gamma^{\rm odd ~(ev)}=\frac12 \kappa\gamma \sum_{q\in X_{\rm odd ~(ev)}} \sin (2\pi q/N)~ (b^+_q b^+_{-q}+b^-_{-q} b^-_q),
\ee

\be
H_h^{\rm odd ~(ev)}=h \sum_{q\in X_{\rm odd ~(ev)}} b^+_q b^-_q -Nh/2.
\ee

\subsection*{A.4~~~ Bogolyubov transformation}

Define the following quantities
\be
 \Gamma_q\equiv \gamma \kappa \sin (2\pi q/N), ~~~ \varepsilon_q\equiv h-\kappa\cos(2\pi q/N),~~~E_q\equiv\sqrt{\varepsilon_q^2+\Gamma_q^2},
\ee

\be
 \theta_q\equiv - \arcsin(\Gamma_q/E_q).
\ee
The Bogolyubov transformation reads
\be
c_q^-=\cos\frac{\theta_q}2 ~b_q^- - \sin \frac{\theta_q}2 ~b_{-q}^+.
\ee
The odd and even parts of the Hamiltonian take the form

\be
H^{\rm odd ~(ev)}= \sum_{q\in X_{\rm odd~(ev)}}~ E_q (c^+_q c^-_q-\frac12).
\ee
This completes the diagonalization.

\subsection*{A.5~~~ Eigenstates}
Let us first construct the (pseudo)vacuum states with respect to the annihilation operators $c_q^-,$ i.e. the states $|{\rm vac}\ra_{\rm odd},~|{\rm vac\ra_{\rm ev}}$ which satisfy
\be
c_q^-|{\rm vac}\ra_{\rm odd~(ev)}=0~~~\forall~q\in X_{\rm odd~(ev)}.
\ee
This states may be written as
\be
|{\rm vac}\ra_{\rm odd}=\aleph_{\rm odd} c_{-N/2+1}^-c_{-N/2+2}^-...c_{N/2}^-|\uparrow\uparrow...\uparrow\ra,
\ee
\be
|{\rm vac}\ra_{\rm ev}=\aleph_{\rm ev} c_{-N/2+1/2}^-c_{-N/2+3/2}^-...c_{N/2-1/2}^-|\uparrow\uparrow...\uparrow\ra,
\ee
where $\aleph_{\rm odd}$ and $\aleph_{\rm ev}$ are normalization constants. This constants are finite, which may be seen from the equalities
\be
|\la\downarrow\downarrow...\downarrow |c_{-N/2+1}^-c_{-N/2+2}^-...c_{N/2}^-|\uparrow\uparrow...\uparrow\ra|=\prod_{q\in X_{\rm odd}} \cos \frac{\theta_q}2\neq 0,
\ee
\be
|\la\downarrow\downarrow...\downarrow |c_{-N/2+1/2}^-c_{-N/2+3/2}^-...c_{N/2-1/2}^-|\uparrow\uparrow...\uparrow\ra|=\prod_{q\in X_{\rm ev}} \cos \frac{\theta_q}2\neq 0.
\ee
Note that $|{\rm vac}\ra_{\rm ev}$ is indeed an eigenstate of the Hamiltonian, while $|{\rm vac}\ra_{\rm odd}$ is not.

All the eigenstates of the Hamiltonian are obtained from the vacuum states by applying the creation operators $c_q^+$. To create the odd number of fermions one should use $ q\in X_{\rm odd} $ and $|{\rm vac}\ra_{\rm odd}$, while to create the even number of fermions one should use  $q\in X_{\rm ev}$ and $|{\rm vac}\ra_{\rm ev}.$

Evidently one can enumerate all the eigenstates of the Hamiltonian by the multiindexes
\be
Q_M \equiv \{q_1,q_2,...,q_M\},~~~0 \leq M \leq N,
\ee
with the ordering $q_1<q_2<...<q_M.$ Then the eigenstate with $M$ fermions read
\be
|Q_M\ra\equiv b^+_{q_1}b^+_{q_2}...b^+_{q_M}|{\rm vac}\ra_{\rm odd (ev)}
\ee
with $q_1,q_2,...,q_M\in X_{\rm odd~(ev)}$  when  $M$ is odd (even).

For our purposes we need only the matrix elements between the states with the same parity, therefore we use the notation $|{\rm vac}\ra$ without subscripts in what follows.

\section*{Appendix B: calculation of $p^z(t)$}

Here we outline the basic steps in obtaining the main result of section 5 -- the formula for $p^z(t)$ in case of the infinite initial temperature of the environment, i.e. eqs.(\ref{pz(t) at beta=0}),(\ref{pz(t) at beta=0, definitions}).


The polarizations which enter the general expression (\ref{p(t)}) now should be written as
\be
p^\alpha_{Q_M \tilde Q_{\tilde M}}=\la \tilde Q_{\tilde M}|\sigma_1^\alpha|Q_M \ra.
\ee
In particular,
\be
p^z_{Q_M \tilde Q_{\tilde M}}=\la \tilde Q_{\tilde M}|2a_1^+a_1^- - 1|Q_M \ra,
\ee
\be
p^x_{Q_M \tilde Q_{\tilde M}}=\la \tilde Q_{\tilde M}|a_1^+ + a_1^-|Q_M \ra,~ p^y_{Q_M \tilde Q_{\tilde M}}=i\la \tilde Q_{\tilde M}|a_1^- - a_1^+|Q_M \ra.
\ee
Evidently, $p^z_{Q_M \tilde Q_{\tilde M}} \neq 0$ only when $M-\tilde M$ is even, while  $p^{x,y}_{Q_M \tilde Q_{\tilde M}} \neq 0$ only when $M-\tilde M$ is odd. Taking this into account one obtains from eq.(\ref{p(t) at beta=0}) 
\be\label{pz(t) in XY}
p^z(t)=2^{-N} p_0^z \sum_{Q_M, \tilde Q_{\tilde M}} |p_{Q_M \tilde Q_{\tilde M}}^z|^2 e^{-i (E(Q_M)-E(\tilde Q_{\tilde M}))t},
\ee
where $E(Q_M)\equiv \sum\limits_{q\in Q_M} E_q.$

The polarizations $p^z_{Q_M \tilde Q_{\tilde M}}$ are nonzero in three cases:
\begin{enumerate}
\item
\be
p^z_{Q_M Q_M}=\frac1N\sum_{p\in X_M}\eta(Q_M,p)\cos\theta_p,
\ee
where $X_M=X_{\rm odd~(ev)}$ if $M$ is odd (even), and $\eta(Q_M,p)=1$ if $p\in Q_M$ and $-1$ otherwise.

\item
\be\nonumber
Q_M=K_{M-1}\cup\{q\},~~~\tilde Q_M=K_{M-1}\cup\{\tilde q\}, ~~~ q,\tilde q \notin K_{M-1},~~q \neq \tilde q:
\ee
\be
|p^z_{Q_M \tilde Q_M}|=\frac2N \cos\frac{\theta_q+\theta_{\tilde q}}2,
\ee

\item
\be\nonumber
Q_M=\tilde Q_{M-2}\cup\{k\}\cup\{\tilde k\},~~~k,\tilde k \notin \tilde Q_{M-2},~~k \neq \tilde k:
\ee
\be
|p^z_{Q_M \tilde Q_{M-2}}|=|p^z_{\tilde Q_{M-2} Q_M}|=\frac2N |\sin\frac{\theta_k-\theta_{\tilde k}}2|
\ee
\end{enumerate}
Using the above formulae one obtains eqs.(\ref{pz(t) at beta=0}),(\ref{pz(t) at beta=0, definitions}) from eq.(\ref{pz(t) in XY}) after straightforward (although somewhat bulky) calculations.


\begin{thebibliography}{9}


\bibitem{Lychkovskiy} Lychkovskiy O
 2010 {\it Phys. Rev. E} {\bf 82} 011123

\bibitem{Izergin} Izergin A G, Kapitonov V S and Kitanin N A 2000 {\it J. Math. Sci.} {\bf 100} 2120

\bibitem{Pasquale} De Pasquale A and Facchi P 2009 {\it Phys. Rev. A} {\bf 80} 032102

\bibitem{Paz and Zurek} Paz J P and Zurek W H 1999 {\it Phys. Rev. Lett.} {\bf 82} 5181

\bibitem{Gogolin} Gogolin C 2010 {\it Phys. Rev. E} {\bf 81} 051127

\bibitem{Niemeijer}  Niemeijer Th 1967 {\it Physica} {\bf 36} 377

\bibitem{Mossel}  Mossel J and Caux J-S 2010
{\it New J. Phys.} {\bf 12} 055028

\bibitem{Lieb} Lieb E, Schultz T and Mattis D 1961 {\it Ann. Phys.} {\bf 16} 407

\end{thebibliography}
\end{document}